\documentclass[aps,prl,twocolumn, showpacs,
  superscriptaddress]{revtex4}

\usepackage{graphicx} \usepackage{color} \usepackage{natbib}
\usepackage{epsfig}

\usepackage{amsmath,amssymb,amsfonts}

\begin{document}

 \title{Prompt merger collapse and the maximum mass of neutron stars}

\author{A.~Bauswein} \affiliation{Max-Planck-Institut f\"ur
  Astrophysik, Karl-Schwarzschild-Str.~1, D-85748 Garching, Germany}
\affiliation{Department of Physics, Aristotle University of
  Thessaloniki, GR-54124 Thessaloniki, Greece}

\author{T.~W.~Baumgarte} \affiliation{Max-Planck-Institut f\"ur
  Astrophysik, Karl-Schwarzschild-Str.~1, D-85748 Garching, Germany} 
  \affiliation{Department of Physics and Astronomy, Bowdoin College, Brunswick, ME 04011, USA}

\author{H.-T.~Janka} \affiliation{Max-Planck-Institut f\"ur
  Astrophysik, Karl-Schwarzschild-Str.~1, D-85748 Garching, Germany}

\date{\today} 

\begin{abstract}
We perform hydrodynamical simulations of neutron-star mergers for a large sample of temperature-dependent, nuclear equations of state, and determine the threshold mass above which the merger remnant promptly collapses to form a black hole.  We find that, depending on the equation of state, the threshold mass is larger than the maximum mass of a nonrotating star in isolation by between 30 and 70 per cent.   Our simulations also show that the ratio between the threshold mass and maximum mass is tightly correlated with the compactness of the nonrotating maximum-mass configuration.  We speculate on how this relation can be used to derive constraints on neutron-star properties from future observations.

   \end{abstract}


   \maketitle

{\it Introduction:} Merging neutron stars (NSs) are among the most promising sources of gravitational radiation for the new generation of gravitational wave (GW) interferometers.
Detection rates for Advanced LIGO~\cite{2010CQGra..27h4006H} and Advanced Virgo~\cite{2006CQGra..23S.635A} have been estimated to be between 0.4 and 400 events per year~\cite{2010CQGra..27q3001A}.   The merger may result either in a black hole (BH) with a hot accretion torus, or a massive, hot, differentially rotating NS.  Compact binary mergers were also suggested as the central engines of short gamma-ray bursts (GRBs)~\cite{1984SvAL...10..177B,1989Natur.340..126E}. Material that becomes gravitationally unbound during the coalescence may undergo rapid neutron-capture nucleosynthesis and contribute to the galactic enrichment by heavy, neutron-rich elements~\cite{1977ApJ...213..225L,1989Natur.340..126E}. The heat release by the radioactive decay of the nucleosynthesis products may also power electromagnetic counterparts~\cite{1998ApJ...507L..59L,2005astro.ph.10256K,2010MNRAS.406.2650M}, which are already being searched for \cite{2013arXiv1306.3960B,2013arXiv1306.4971T}.

The dynamics and observable signatures of NS merger depend on the binary masses $M_{1,2}$ and the equation of state (EoS)~\cite{2005PhRvL..94t1101S,2005PhRvD..71h4021S,2006PhRvD..73f4027S,2007A&A...467..395O,2008PhRvD..77b4006A,2008PhRvD..78b4012L,2008PhRvD..78h4033B,2009PhRvD..80f4037K,2011PhRvD..83l4008H,2011PhRvL.107e1102S,2012PhRvD..85j4030B,2012PhRvD..86f3001B,2013MNRAS.430.2585R,2013arXiv1302.6530B} (see also~\cite{2010CQGra..27k4002D,BaumgarteShapiro,2012LRR....15....8F} for reviews).  At nuclear densities, the EoS is not completely known (see, e.g., \cite{2012ARNPS..62..485L}) but plays a crucial role in determining the immediate outcome of coalescence. For sufficiently low-mass binaries the merger results in a stable NS. For more massive binaries the remnant will ultimately form a BH. In the {\em delayed collapse} scenario, the two stars form a single, differentially rotating merger remnant that is temporarily supported against gravitational collapse by centrifugal and thermal effects \cite{2000ApJ...528L..29B,2013arXiv1306.4034K}.  Viscous processes, radiation of GWs and emission of neutrinos redistribute and reduce the remnant's angular momentum and energy, prompting a delayed collapse on a secular timescale.   Alternatively, the merger may lead to an immediate, {\em prompt collapse} on a dynamical timescale.  Such a collapse is triggered for more massive binaries, whose total mass $M_{\mathrm{tot}}=M_1+M_2$ cannot be stabilized.  For a given EoS one can thus define a threshold binary mass $M_{\mathrm{thres}}$ that separates the two scenarios of prompt and delayed collapse.  The former occurs for $M_{\mathrm{tot}}>M_{\mathrm{thres}}$, while a dynamically stable remnant is formed for $M_{\mathrm{tot}}<M_{\mathrm{thres}}$.

It is intuitive to assume that $M_{\mathrm{thres}}$ scales with the maximum mass $M_{\mathrm{max}}$ of isolated, nonrotating NSs~\cite{2011PhRvD..83l4008H},
\begin{equation}\label{eq:k}
M_{\mathrm{thres}}=k\cdot M_{\mathrm{max}}.
\end{equation}
Here $M_{\mathrm{max}}$ is determined by the EoS and can be found by integrating the 
Tolman-Oppenheimer-Volkoff (TOV) equations (equations of relativistic hydrostatic equilibrium)~\cite{1939PhRv...55..364T,1939PhRv...55..374O}.   The coefficient $k$ also depends on the EoS, or equivalently on NS properties~\cite{2005PhRvL..94t1101S,2005PhRvD..71h4021S,2006PhRvD..73f4027S,2011PhRvD..83l4008H}.  

In this paper we adopt a large set of temperature-dependent, nuclear EoSs in numerical simulations of binary neutron-star mergers to examine the dependence of $k$ on the EoS, and to establish a relation between $M_{\mathrm{thres}}$ and $M_{\mathrm{max}}$.  We focus on equal-mass binaries, but also comment on asymmetric systems below.  We find that $k$ is tightly correlated with the compactness $C_{\mathrm{max}}=(G M_{\mathrm{max}})/(c^2 R_{\mathrm{max}})$ of the maximum-mass TOV configuration ($G$ is the gravitational constant and $c$ the speed of light).  We provide a simple, analytical model to motivate such a correlation, and discuss how our results can be used to constrain NS properties, in particular $M_{\mathrm{max}}$, from future observations. For a given EoS our findings predict which binary systems undergo prompt or delayed collapse upon merger with corresponding consequences for the post-merger GW signal, the mass ejection during coalescence and the particular conditions for launching a collimated outflow favorable for a GRB (e.g.~torus properties and baryon loading of the environment).

{\it Method:} We perform numerical simulations of NS mergers to determine the EoS dependence of $M_{\mathrm{thres}}$, using a 3D relativistic smoothed particle hydrodynamics (SPH) code that employs the conformal flatness approximation of Einstein's field equations and includes a GW backreaction scheme to account for energy and angular momentum losses due to GW emission (see~\cite{2002PhRvD..65j3005O,2007A&A...467..395O,2010PhRvD..82h4043B} for details of the code).  Our study considers 12 microphysical, fully temperature-dependent EoSs with maximum masses in the range of 1.95 to 2.79~$M_{\odot}$, which is compatible with the observation of a $1.97\pm 0.04~M_{\odot}$ pulsar~\cite{2010Natur.467.1081D} (see Tab.~\ref{tab:models}).  With the exception of the IUF EoS, these EoSs are also consistent with the detection of a NS with a mass of $2.01\pm 0.04~M_{\odot}$~\cite{Antoniadis26042013}. The radii $R_{\mathrm{max}}$ of the maximum-mass configurations vary between 10.32 and 13.43~km (see also~\cite{2012PhRvD..86f3001B} for the mass-radius relations of most EoSs considered here). The EoSs are chosen without any selection procedure and cover approximately the full range of high-density models regarding their stellar properties. As initial conditions we set up cold NSs in neutrinoless beta-equilibrium on a quasi-equilibrium orbit a few revolutions before merging. We assume irrotational stars since tidal locking is unlikely~\cite{1992ApJ...400..175B,1992ApJ...398..234K} and the orbital period is short compared to possible stellar rotation.  Unless stated otherwise we use a resolution of about 340,000 SPH particles.

For each EoS we determine $M_{\mathrm{thres}}$ by performing simulations of binaries with different values of $M_{\mathrm{tot}}$, which is defined as the binary's total gravitational mass at infinitely large binary separation.  We focus on equal-mass binaries here and increase $M_{\mathrm{tot}}$ in increments of 0.1~$M_{\odot}$.  We identify  $M_{\mathrm{stab}}$ with the mass of the most massive binary in our sample with dynamically stable remnant, i.e.~the most massive system that results in a delayed collapse. We similarly identify $M_{\mathrm{unstab}}$ with the mass of the least massive binary whose merger triggers prompt collapse. We then estimate $M_{\mathrm{thres}} = (M_{\mathrm{stab}}+M_{\mathrm{unstab}})/2 \pm 0.05 M_{\odot}$.

\begin{table}
\caption{\label{tab:models} Sample of termperature-dependent, nuclear EoSs used in this study.  Here $M_{\mathrm{max}}$, $R_{\mathrm{max}}$, $C_{\mathrm{max}}$, and $\rho_{\mathrm{c}}$ are the gravitational mass, areal radius, compactness, and central energy density of the maximum-mass TOV configurations. We list $\rho_{\mathrm{c}}$ in units of the nuclear saturation density $\rho_0=2.7\times 10^{14}~\mathrm{g/cm^3}$. $R_{1.6}$ is the areal radius of 1.6~$M_{\odot}$ NSs. $M_{\mathrm{thres}}$ denotes the total binary mass that separates prompt from delayed collapse (see text). $f_{\mathrm{peak}}^{\mathrm{stab}}$ is the dominant GW frequency in the post-merger phase of the binary with $M_{\mathrm{tot}}=M_{\mathrm{stab}}$, the most massive binary configuration of our sample that does not collapse promptly.}
\begin{ruledtabular}
\begin{tabular}{|l|l|l|l|l|l|l|l|}
EoS     & $M_{\mathrm{max}}$ & $R_{\mathrm{max}}$ & $C_{\mathrm{max}}$ & $R_{1.6}$ & $M_{\mathrm{thres}}$ & $\rho_{\mathrm{c}}/\rho_0$ & $f_{\mathrm{peak}}^{\mathrm{stab}}$\\
                        & $[M_{\odot}]$ & [km] & & [km] & $[M_{\odot}]$ &  & [kHz] \\ \hline
NL3~\cite{1997PhRvC..55..540L,2010NuPhA.837..210H}   & 2.79 & 13.43 & 0.307 & 14.81 & 3.85 & 5.6  & 2.78 \\
GS1~\cite{2011PhRvC..83c5802S}                       & 2.75 & 13.27 & 0.306 & 14.79 & 3.85 & 5.7  & 2.81 \\
LS375~\cite{1991NuPhA.535..331L}                     & 2.71 & 12.34 & 0.325 & 13.71 & 3.65 & 6.5  & 3.05 \\
DD2~\cite{2010PhRvC..81a5803T,2010NuPhA.837..210H}   & 2.42 & 11.90 & 0.300 & 13.26 & 3.35 & 7.2  & 3.06 \\
Shen~\cite{1998NuPhA.637..435S}                      & 2.22 & 13.12 & 0.250 & 14.46 & 3.45 & 6.7  & 2.85 \\
TM1~\cite{1994NuPhA.579..557S,2012ApJ...748...70H}   & 2.21 & 12.57 & 0.260 & 14.36 & 3.45 & 6.7  & 2.91 \\
SFHX~\cite{2012arXiv1207.2184S}                      & 2.13 & 10.76 & 0.292 & 11.98 & 3.05 & 8.9  & 3.52 \\
GS2~\cite{2011arXiv1103.5174S}                       & 2.09 & 11.78 & 0.262 & 13.31 & 3.25 & 7.6  & 3.19 \\
SFHO~\cite{2012arXiv1207.2184S}                      & 2.06 & 10.32 & 0.294 & 11.76 & 2.95 & 9.8  & 3.67 \\
LS220~\cite{1991NuPhA.535..331L}                     & 2.04 & 10.62 & 0.284 & 12.43 & 3.05 & 9.4  & 3.52 \\
TMA~\cite{1995NuPhA.588..357T,2012ApJ...748...70H}   & 2.02 & 12.09 & 0.247 & 13.73 & 3.25 & 7.2  & 2.96 \\
IUF~\cite{2010PhRvC..82e5803F,2010NuPhA.837..210H}   & 1.95 & 11.31 & 0.255 & 12.57 & 3.05 & 8.1  & 3.31 \\
\end{tabular}
\end{ruledtabular} 
\end{table}

Since thermal pressure has an important effect on the collapse behavior (see, e.g.,~\cite{2010PhRvD..82h4043B,2012PhRvD..86f4032P,2013arXiv1306.4034K}), we have only considered fully temperature-dependent EoSs in this study.  Many other simulations instead supplement a barotropic, zero-temperature EoS  with a thermal ideal-gas component in order to approximate finite-temperature effects~\cite{2005PhRvL..94t1101S,2005PhRvD..71h4021S,2006PhRvD..73f4027S,2009PhRvD..80f4037K,2010PhRvD..82h4043B,2010CQGra..27k4002D,2011PhRvD..83l4008H,2012PhRvD..86f3001B}. We have found that in such a ``hybrid" treatment the threshold mass $M_{\mathrm{thres}}$ depends strongly on the ideal-gas index $\Gamma_{\mathrm{th}}$.  Since $\Gamma_{\mathrm{th}}$ is neither unambiguously defined nor constant~\cite{2010PhRvD..82h4043B}, fully temperature-dependent EoSs will provide more reliable values for $M_{\mathrm{thres}}$ than a hybrid treatment.   

In order to calibrate the error introduced by the conformal flatness
approximation we reproduced the fully relativistic simulations
of~\cite{2011PhRvD..83l4008H} and found the same collapse behavior in
all but one case, for which we obtained a small shift in
$M_{\mathrm{thres}}$ \footnote{We compared models for the ALF2, SLy4
  and APR4 EoSs, using a ``hybrid" treatment for thermal effects, and
found the same behavior as in~\cite{2011PhRvD..83l4008H} for all
  cases except for APR4, for which we obtained prompt collapse for a
  1.42-1.42~$M_{\odot}$ binary instead of a 1.4-1.4~$M_{\odot}$
  system.}.  We conclude that the effects of the conformal flatness
approximation on our results are small.  We verified that our
resolution with SPH particles is sufficient by reproducing our
findings for the DD2 EoS with both 731,000 and 1,202,000 SPH
particles. Finally, we reran our simulations for the DD2 EoS starting
with different initial binary separations (leading to 2.5, 3.5 and 4.5
orbits before merging) to confirm that this separation does not affect
our results.

\begin{figure}
\includegraphics[width=8.7cm]{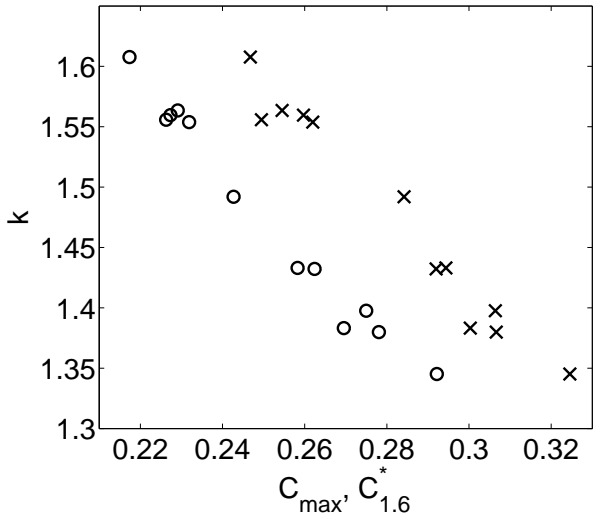}
\caption{\label{fig:kcmax}Coefficient $k$ (eq.~(\ref{eq:k})) as a function of  $C_{\mathrm{max}}= G M_{\mathrm{max}}/(c^2 R_{\mathrm{max}})$ (crosses) and ${C}^*_{1.6}=G M_{\mathrm{max}}/(c^2 R_{1.6})$ (circles).}
\end{figure}

{\it Results:} The EoS dependence of $M_{\mathrm{thres}}$ and $k$ can
be expressed by the stellar parameters of nonrotating NSs, which are
uniquely determined by the EoS and thus characterize a given EoS. Our
survey reveals that $k$ scales very well with the compactness
$C_{\mathrm{max}}=(GM_{\mathrm{max}})/(c^2R_{\mathrm{max}})$ of the
maximum-mass configuration of nonrotating NSs (Fig.~\ref{fig:kcmax}).
We find a similarly tight relation when $k$ is expressed as a function
of ${C}^*_{1.6}=(GM_{\mathrm{max}})/(c^2R_{1.6})$, where $R_{1.6}$ is
the radius of a 1.6~$M_{\odot}$ NS (see Fig.~\ref{fig:kcmax}).   Since
$R_{1.6}$ may be more easily determined than $R_{\mathrm{max}}$,
both by future
observations~\cite{2012ARNPS..62..485L,2013arXiv1306.4065R,2012PhRvL.108a1101B,2012PhRvD..86f3001B}
and theoretical considerations~\cite{2010PhRvL.105p1102H}, $ C^*_{1.6}$ might be a more useful quantity than $C_{\mathrm{max}}$.

\begin{figure}
\includegraphics[width=8.9cm]{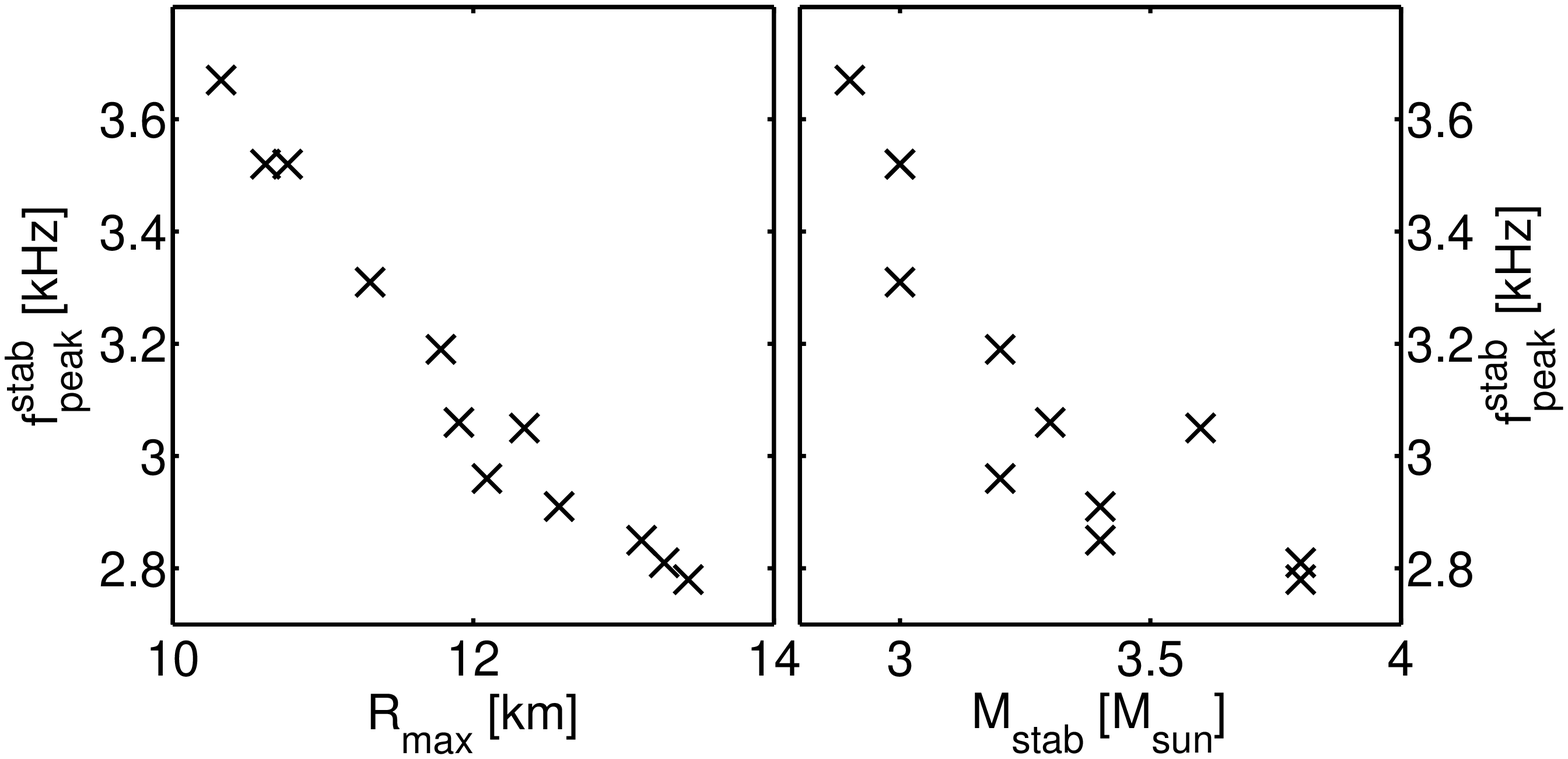}
\caption{\label{fig:fpeak} Dominant GW frequency $f_{\mathrm{peak}}^{\mathrm{stab}}$ of the post-merger phase for different NS EoSs as function of radius $R_{\mathrm{max}}$ (left panel) and $M_{\mathrm{stab}}$ (right panel). For each EoSs, $f_{\mathrm{peak}}^{\mathrm{stab}}$ is the frequency $f_{\mathrm{peak}}$ for the most massive dynamically stable binary ($M_{\mathrm{tot}}=M_{\mathrm{stab}}$).}
\end{figure}

As can be seen in Fig.~\ref{fig:kcmax}, $k$ is a nearly linear
function of ${C}^*_{1.6}$ in the regime of interest.  The maximum
residual from the linear fit $k=j\cdot {C}^*_{1.6}+a$ with $j=-3.606$
and $a=2.380$ is only 0.025 \footnote{We found that similar quasi-linear
  relations also hold for $C_{\mathrm{max}}$ and for any other
  ${C}^*=(G M_{\mathrm{max}})/(c^2R_{\mathrm{NS}})$, in which NS radii
  $R_{\mathrm{NS}}$ different from those of 1.6~$M_{\odot}$ are used
  as EoS characterization, e.g. ${C}^*_{1.4}=(GM_{\mathrm{max}})/(c^2R_{1.4})$.}. By fixing $R_{1.6}$ or $R_{\mathrm{max}}$ (see also the discussion of Fig.~\ref{fig:fpeak}), $M_{\mathrm{thres}}$ becomes a quadratic function of $M_{\mathrm{max}}$ only. Considering the maximum deviation of $k$ from the fit implies that $M_{\mathrm{thres}}$ can be converted to $M_{\mathrm{max}}$ with a precision of a few per cent for a fixed $R_{1.6}$. An uncertainty of, for instance, 0.5~km in $R_{1.6}$ would add another $\sim 5$~per cent error. The actual error may be smaller because the deviation of $k$ from the fit includes the intrinstic scatter among different EoSs but also an artificial contribution from the finite sampling of $M_{\mathrm{tot}}$ values.

\begin{figure}
\includegraphics[width=8.9cm]{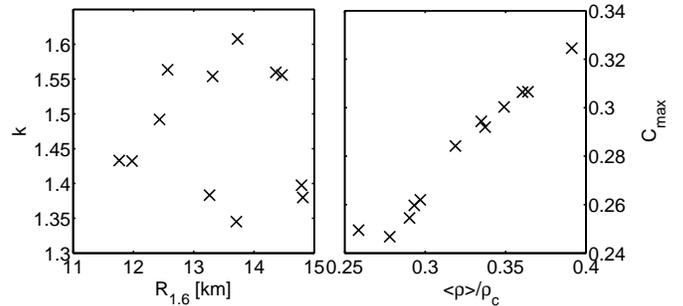}
\caption{\label{fig:doublekcmax} 
{\em Left panel:} Coefficient $k$ (eq.~(\ref{eq:k})) versus radius $R_{1.6}$ of a 1.6~$M_{\odot}$ NSs.  {\em Right panel:} Compactness $C_{\mathrm{max}}$ as a function of the EoS's stiffness expressed by the ratio of the average density $\langle\rho\rangle=3 M_{\mathrm{max}}/(4\pi R_{\mathrm{max}}^3)$ and central energy density $\rho_{\mathrm{c}}$.}
\end{figure}

We compared our findings with those of \cite{2011PhRvD..83l4008H}, where six barotropic EoSs with a hybrid treatment of finite-temperature effects were adopted and an approximate relation between $k$ and the radius $R_{1.4}$ of a 1.4~$M_{\odot}$ NS was suggested.  Testing this relationship with our extended set of temperature-dependent EoSs results in a distribution with rather wide scattering instead of a tight correlation (see Fig.~\ref{fig:doublekcmax}, left panel, and Tab.~\ref{tab:models}; $R_{1.6}$ is very similar to $R_{1.4}$). However, using the numerical data of~\cite{2011PhRvD..83l4008H} and expressing $k$ as a function of $C^*_{1.6}$ or $C_{\mathrm{max}}$ rather than $R_{1.4}$, we found a tight correlation, as for our results. Therefore, we suspect that the approximate scaling with $R_{1.4}$ suggested in~\cite{2011PhRvD..83l4008H} is a selection effect due to the limited number of EoSs used therein \footnote{The hybrid treatment of thermal effects employed in~\cite{2011PhRvD..83l4008H}, with a relatively small value of $\Gamma_{\mathrm{th}}=1.357$, is likely to underestimate thermal pressure support~\cite{2010PhRvD..82h4043B}. This should be taken into account in a direct comparison of the $k$ values.}.

The compactness $C_{\mathrm{max}}$ is a measure of the EoS's stiffness at high densities (Fig.~\ref{fig:doublekcmax}, right panel; see also~\cite{1997ApJ...488..799K,2012ARNPS..62..485L}), where we characterize the stiffness by the ratio of the mean density, $\langle\rho\rangle=3 M_{\mathrm{max}}/(4\pi R_{\mathrm{max}}^3)$, to the central density $\rho_\mathrm{c}$ (i.e.~the inverse central condensation). A tight correlation between $k$ and $C_{\mathrm{max}}$ thus implies that $k$ depends predominantly on the stiffness of the EoS.  This dependence can be motivated qualitatively with the help of a simple Newtonian model. As suggested in \cite{2003ApJ...583..410L}, a rough estimate of the fractional increase in the maximum mass, $\delta M/M_{\mathrm{max}}$, is given by $3 \, T/|W|$, so that $k \approx 1 + 3\, T/|W|$.  Here $T$ is the rotational kinetic energy and $W$ the potential energy.   We compute $T = J^2/(2I)$, where $I$ is the remnant's moment of inertia, from the angular momentum $J$ that the binary carries at the instant of merging.  Approximating the merging of an equal-mass binary in circular orbit to occur when the binary separation is twice the radius of each individual (spherical) star, $R_\star$, and assuming that the progenitors' masses are concentrated at their centers, we find $J^2 \approx G M^3_{\rm tot} R_\star/8$.  Neglecting mass loss as well as deviations from spherical symmetry, and assuming that the merger remnant forms a polytrope with polytropic index $n$, we have $W = -3 G/(5-n) \, M_{\rm tot}^2 / R$, where $R$ is the radius of the remnant, and $I = 2 \kappa_n M_{\rm tot} R^2/5$.  Here the coefficients $\kappa_n$ depend on $n$ only and are tabulated in \cite{1993ApJS...88..205L}. The EoS's stiffness as well as $\kappa_n$ increases with decreasing $n$.  Using the polytropic mass-radius relationship for the merging NSs and merger remnant we also have $R_\star/R = 2^{(n-1)/(3-n)}$.  Collecting terms we now obtain $k\approx 1 + 5(5-n)\,2^{(n-1)/(3-n)}/(32 \kappa_n) $.  While this crude approximation overestimates the deviation of $k$ from unity by about a factor of two, it correctly predicts two important qualitative features of our numerical results: It suggests that $k$ depends predominantly on the EoS's stiffness (since for Newtonian polytropes the stiffness $\langle\rho\rangle/\rho_{\mathrm{c}}$ depends on $n$ only), and it shows that $k$ decreases with increasing stiffness (which can be seen by inserting values for $n$ and $\kappa_n$). Loosely speaking, a binary with a stiffer EoS (i.e.~a larger $\langle\rho\rangle/\rho_{\mathrm{c}}$) has less angular momentum when merging and its remnant has a larger moment of inertia. These effects combine to decrease $T/|W|$, thereby decreasing $k$.

For the EoSs in our sample we also observe a tight correlation between  $R_{\mathrm{max}}$ and $R_{1.6}$, which implies a close relation between $C_{\mathrm{max}}$ and ${C}^*_{1.6}$.

{\it Observational constraints on the maximum NS mass:} The findings of this study may help to place limits on the maximum mass $M_{\mathrm{max}}$ of NSs in the case that future observations, e.g.~GW detections, provide an estimate of $M_{\mathrm{thres}}$ (cf.~\cite{2005PhRvL..94t1101S}). We assume that delayed and prompt collapse can be distinguished from the presence or absence of GW emission in the 2-4~kHz range produced by the oscillations of the merger remnant, and that the binary mass of the merger can be inferred from the preceeding GW inspiral signal, which thus sets a bound on $M_{\mathrm{thres}}$.  Depending on the nature of available observations, this information could be used in different ways.  In the following we discuss three speculative possibilities.  

We first assume that a number of detections of NS mergers have been
made, and that observations of both prompt and delayed collapses
bracket $M_{\mathrm{thres}}$ to a certain accuracy.  If $R_{1.6}$
  (or the radius of a NS of any other mass~[65]) is independently known to some accuracy, either from future GW measurements (e.g.~\cite{2012PhRvL.108a1101B,2012PhRvD..86f3001B,2013arXiv1306.4065R}) or astronomical observations~\cite{2012ARNPS..62..485L}, then the relation shown in Fig.~\ref{fig:kcmax} provides an estimate for the maximum mass of nonrotating NSs in isolation.  The accuracy of this estimate depends on the accuracy of $M_{\mathrm{thres}}$ and $R_{1.6}$, of course, but since the scatter in the relation between $k$ and $ C^*_{1.6}$ is quite small, this scatter will contribute only a few per cent of error.

As our second example we point out that even a single observation of a delayed collapse could provide both an upper and a lower bound on $M_{\mathrm{thres}}$.  The lower bound is given immediately by the measured binary mass $M_{\mathrm{tot}}$. An upper bound can be established from the dominant GW frequency $f_{\mathrm{peak}}$ of the post-merger oscillations. To show this, we note that $f_{\mathrm{peak}}$ increases with increasing binary mass for a given EoS~\cite{2006PhRvD..73f4027S,2012PhRvD..86f3001B}.  The measured value of $f_{\mathrm{peak}}$ therefore provides a lower limit for the peak frequency $f_{\mathrm{peak}}^{\mathrm{stab}}$ of a binary with the highest total binary mass $M_{\mathrm{stab}}$ leading to delayed collapse.  In Fig.~\ref{fig:fpeak} we show that $f_{\mathrm{peak}}^{\mathrm{stab}}$ exhibits a tight anti-correlation with $R_{\mathrm{max}}$, and a somewhat looser anti-correlation with $M_{\mathrm{stab}}$, which approximates $M_{\mathrm{thres}}$.  The lower limit on $f_{\mathrm{peak}}^{\mathrm{stab}}$ therefore provides an upper limit on both $M_{\mathrm{thres}}$ and $R_{\rm max}$.   This means that a measurement of $f_{\mathrm{peak}}$ and the binary masses establishes both an upper and a lower limit on $M_{\rm thres}$, as well as an upper limit on $R_{\rm max}$.  These bounds can then be combined to establish a constraint on $M_{\rm max}$.  

In the third scenario we again consider just a single detection, but
this time we assume that prompt collapse has been established
unambiguously.  The measured binary mass $M_{\rm tot}$ then forms an
upper limit for $M_{\mathrm{thres}}$ (as well as a
  lower limit of $M_{\rm tot}/2$).  Without additional information for $R_{\max}$ or $R_{1.6}$, the tightest possible constraint on $M_{\rm max}$ is then $M_{\mathrm{max}} = M_{\rm thres}/k < M_{\rm tot}/k_{\rm min}$, where $k_{\rm min}$ is the smallest conceivable value of $k$.  For concreteness, consider a 1.5-1.5~$M_{\odot}$ binary system that leads to prompt collapse.  Assuming that our EoS sample covers the full range of high-density models, we have $k_{\rm min} \approx 1.3$. Therefore, a single prompt-collapse detection with $M_{\mathrm{tot}}=3~M_{\odot}$ leads to the conclusion that $M_{\mathrm{max}} \leq 3~M_{\odot}/1.3\approx 2.3~M_{\odot}$.

All three of our examples above hinge, of course, on measurements of $M_{\mathrm{tot}}$ and/or $f_{\mathrm{peak}}$.  Measuring the former requires a relatively large signal-to-noise ratio, since, to leading order, the GW phasing during the inspiral depends on the binary's chirp mass rather than the total mass~\cite{1993PhRvD..47.2198F,1994PhRvD..49.2658C,2005PhRvD..71h4008A,2013ApJ...766L..14H,2013arXiv1304.1775T}. Since the dominant post-merger GW frequencies are outside of the most sensitive range of the 
upcoming GW interferometers, detecting these signals will also be possible only for near-by events.

In principle, the precise identification of $M_{\mathrm{thres}}$ is problematic because, as $M_{\mathrm{tot}}$ approaches $M_{\mathrm{thres}}$ from below, the remnant's lifetime becomes increasingly short and the post-merger signal increasingly weak. In practice, however, the lifetime has a steep sensitivity to the total binary mass. For instance, we find that eight out of our 12 simulations with $M_{\mathrm{tot}}=M_{\mathrm{stab}}$ yield remnant lifetimes exceeding 10~ms. This shows that binary systems with masses only slightly below $M_{\mathrm{thres}}$ already result in relatively long-lived remnants.

{\em Discussion:}  As stated before, we have assumed equal-mass binaries in the above simulations.  This assumption is not entirely unjustified, since observations of binary NS systems suggest small mass ratios (see, e.g.,~\cite{2012ARNPS..62..485L} for a review).  However, in order to evaluate the effect of unequal masses we have performed additional simulations with the SFHO, DD2, and NL3 EoSs for binaries with total masses $M_{\mathrm{stab}}$ and $M_{\mathrm{unstab}}$ (as found from the equal-mass simulations), but now with a mass ratio $q\equiv M_1/M_2\approx 0.9$. For the given sampling of $M_{\mathrm{tot}}$ we found that symmetric and asymmetric systems of the same total mass show the same collapse behavior. Generalizing our analytical considerations to non-equal masses, and expanding the result in deviations from symmetry, $\epsilon \equiv q-1$, shows that corrections appear at order $\epsilon^2$. This corroborates our numerical finding that moderate deviations of the mass ratio from unity have a small effect on $k$.

Future work could improve our study in a number of different ways.  Even though we expect that effects of the conformal flatness approximation are small, it would be desirable to perform similar simulations with a fully relativistic treatment. These simulations should also include an even larger sample of temperature-dependent EoSs and should explore the effects of asymmetric binaries.

\begin{acknowledgments}
We thank M. Hempel for providing EoS tables. TWB is grateful for
support from the Alexander-von-Humboldt Foundation and thanks the
Max-Planck-Institut f\"ur Astrophysik for its hospitality.   This work
was supported by DFG grants SFB/TR~7 and EXC~153, by ESF/CompStar, RZG
Garching, as well as NSF grant PHY-1063240 to Bowdoin College. A.B. is a Marie Curie Intra-European Fellow within the 7th European Community
Framework Programme (IEF 331873).
\end{acknowledgments}


\begin{thebibliography}{63}
\expandafter\ifx\csname natexlab\endcsname\relax\def\natexlab#1{#1}\fi
\expandafter\ifx\csname bibnamefont\endcsname\relax
  \def\bibnamefont#1{#1}\fi
\expandafter\ifx\csname bibfnamefont\endcsname\relax
  \def\bibfnamefont#1{#1}\fi
\expandafter\ifx\csname citenamefont\endcsname\relax
  \def\citenamefont#1{#1}\fi
\expandafter\ifx\csname url\endcsname\relax
  \def\url#1{\texttt{#1}}\fi
\expandafter\ifx\csname urlprefix\endcsname\relax\def\urlprefix{URL }\fi
\providecommand{\bibinfo}[2]{#2}
\providecommand{\eprint}[2][]{\url{#2}}

\bibitem[{\citenamefont{{Harry} and {LIGO Scientific
  Collaboration}}(2010)}]{2010CQGra..27h4006H}
\bibinfo{author}{\bibfnamefont{G.~M.} \bibnamefont{{Harry}}} \bibnamefont{and}
  \bibinfo{author}{\bibnamefont{{LIGO Scientific Collaboration}}},
  \bibinfo{journal}{Classical and Quantum Gravity}
  \textbf{\bibinfo{volume}{27}}, \bibinfo{eid}{084006} (\bibinfo{year}{2010}).

\bibitem[{\citenamefont{{Acernese} et~al.}(2006)\citenamefont{{Acernese},
  {Amico}, {Alshourbagy}, {Antonucci}, {Aoudia}, {Avino}, {Babusci},
  {Ballardin}, {Barone}, {Barsotti} et~al.}}]{2006CQGra..23S.635A}
\bibinfo{author}{\bibfnamefont{F.}~\bibnamefont{{Acernese}}},
  \bibinfo{author}{\bibfnamefont{P.}~\bibnamefont{{Amico}}},
  \bibinfo{author}{\bibfnamefont{M.}~\bibnamefont{{Alshourbagy}}},
  \bibinfo{author}{\bibfnamefont{F.}~\bibnamefont{{Antonucci}}},
  \bibinfo{author}{\bibfnamefont{S.}~\bibnamefont{{Aoudia}}},
  \bibinfo{author}{\bibfnamefont{S.}~\bibnamefont{{Avino}}},
  \bibinfo{author}{\bibfnamefont{D.}~\bibnamefont{{Babusci}}},
  \bibinfo{author}{\bibfnamefont{G.}~\bibnamefont{{Ballardin}}},
  \bibinfo{author}{\bibfnamefont{F.}~\bibnamefont{{Barone}}},
  \bibinfo{author}{\bibfnamefont{L.}~\bibnamefont{{Barsotti}}},
  \bibnamefont{et~al.}, \bibinfo{journal}{Classical Quantum Gravity}
  \textbf{\bibinfo{volume}{23}}, \bibinfo{pages}{635} (\bibinfo{year}{2006}).

\bibitem[{\citenamefont{{Abadie} et~al.}(2010)}]{2010CQGra..27q3001A}
\bibinfo{author}{\bibfnamefont{J.}~\bibnamefont{{Abadie}}}
  \bibnamefont{et~al.}, \bibinfo{journal}{Class. Quantum Grav.}
  \textbf{\bibinfo{volume}{27}}, \bibinfo{pages}{173001}
  (\bibinfo{year}{2010}).

\bibitem[{\citenamefont{{Blinnikov} et~al.}(1984)\citenamefont{{Blinnikov},
  {Novikov}, {Perevodchikova}, and {Polnarev}}}]{1984SvAL...10..177B}
\bibinfo{author}{\bibfnamefont{S.~I.} \bibnamefont{{Blinnikov}}},
  \bibinfo{author}{\bibfnamefont{I.~D.} \bibnamefont{{Novikov}}},
  \bibinfo{author}{\bibfnamefont{T.~V.} \bibnamefont{{Perevodchikova}}},
  \bibnamefont{and} \bibinfo{author}{\bibfnamefont{A.~G.}
  \bibnamefont{{Polnarev}}}, \bibinfo{journal}{Soviet Astronomy Letters}
  \textbf{\bibinfo{volume}{10}}, \bibinfo{pages}{177} (\bibinfo{year}{1984}).

\bibitem[{\citenamefont{{Eichler} et~al.}(1989)\citenamefont{{Eichler},
  {Livio}, {Piran}, and {Schramm}}}]{1989Natur.340..126E}
\bibinfo{author}{\bibfnamefont{D.}~\bibnamefont{{Eichler}}},
  \bibinfo{author}{\bibfnamefont{M.}~\bibnamefont{{Livio}}},
  \bibinfo{author}{\bibfnamefont{T.}~\bibnamefont{{Piran}}}, \bibnamefont{and}
  \bibinfo{author}{\bibfnamefont{D.~N.} \bibnamefont{{Schramm}}},
  \bibinfo{journal}{\nat} \textbf{\bibinfo{volume}{340}}, \bibinfo{pages}{126}
  (\bibinfo{year}{1989}).

\bibitem[{\citenamefont{{Lattimer} et~al.}(1977)\citenamefont{{Lattimer},
  {Mackie}, {Ravenhall}, and {Schramm}}}]{1977ApJ...213..225L}
\bibinfo{author}{\bibfnamefont{J.~M.} \bibnamefont{{Lattimer}}},
  \bibinfo{author}{\bibfnamefont{F.}~\bibnamefont{{Mackie}}},
  \bibinfo{author}{\bibfnamefont{D.~G.} \bibnamefont{{Ravenhall}}},
  \bibnamefont{and} \bibinfo{author}{\bibfnamefont{D.~N.}
  \bibnamefont{{Schramm}}}, \bibinfo{journal}{\apj}
  \textbf{\bibinfo{volume}{213}}, \bibinfo{pages}{225} (\bibinfo{year}{1977}).

\bibitem[{\citenamefont{{Li} and {Paczy{\'n}ski}}(1998)}]{1998ApJ...507L..59L}
\bibinfo{author}{\bibfnamefont{L.-X.} \bibnamefont{{Li}}} \bibnamefont{and}
  \bibinfo{author}{\bibfnamefont{B.}~\bibnamefont{{Paczy{\'n}ski}}},
  \bibinfo{journal}{Astrophys. J. Lett.} \textbf{\bibinfo{volume}{507}},
  \bibinfo{pages}{L59} (\bibinfo{year}{1998}).

\bibitem[{\citenamefont{{Kulkarni}}(2005)}]{2005astro.ph.10256K}
\bibinfo{author}{\bibfnamefont{S.~R.} \bibnamefont{{Kulkarni}}},
  \bibinfo{journal}{ArXiv e-prints}  (\bibinfo{year}{2005}),
  \eprint{astro-ph/0510256}.

\bibitem[{\citenamefont{{Metzger} et~al.}(2010)\citenamefont{{Metzger},
  {Mart{\'{\i}}nez-Pinedo}, {Darbha}, {Quataert}, {Arcones}, {Kasen}, {Thomas},
  {Nugent}, {Panov}, and {Zinner}}}]{2010MNRAS.406.2650M}
\bibinfo{author}{\bibfnamefont{B.~D.} \bibnamefont{{Metzger}}},
  \bibinfo{author}{\bibfnamefont{G.}~\bibnamefont{{Mart{\'{\i}}nez-Pinedo}}},
  \bibinfo{author}{\bibfnamefont{S.}~\bibnamefont{{Darbha}}},
  \bibinfo{author}{\bibfnamefont{E.}~\bibnamefont{{Quataert}}},
  \bibinfo{author}{\bibfnamefont{A.}~\bibnamefont{{Arcones}}},
  \bibinfo{author}{\bibfnamefont{D.}~\bibnamefont{{Kasen}}},
  \bibinfo{author}{\bibfnamefont{R.}~\bibnamefont{{Thomas}}},
  \bibinfo{author}{\bibfnamefont{P.}~\bibnamefont{{Nugent}}},
  \bibinfo{author}{\bibfnamefont{I.~V.} \bibnamefont{{Panov}}},
  \bibnamefont{and} \bibinfo{author}{\bibfnamefont{N.~T.}
  \bibnamefont{{Zinner}}}, \bibinfo{journal}{Mon. Not. R. Astron. Soc.}
  \textbf{\bibinfo{volume}{406}}, \bibinfo{pages}{2650} (\bibinfo{year}{2010}).

\bibitem[{\citenamefont{{Berger} et~al.}(2013)\citenamefont{{Berger}, {Fong},
  and {Chornock}}}]{2013arXiv1306.3960B}
\bibinfo{author}{\bibfnamefont{E.}~\bibnamefont{{Berger}}},
  \bibinfo{author}{\bibfnamefont{W.}~\bibnamefont{{Fong}}}, \bibnamefont{and}
  \bibinfo{author}{\bibfnamefont{R.}~\bibnamefont{{Chornock}}},
  \bibinfo{journal}{ArXiv e-prints}  (\bibinfo{year}{2013}),
  \eprint{1306.3960}.

\bibitem[{\citenamefont{{Tanvir} et~al.}(2013)\citenamefont{{Tanvir}, {Levan},
  {Fruchter}, {Hjorth}, {Wiersema}, {Tunnicliffe}, and {de Ugarte
  Postigo}}}]{2013arXiv1306.4971T}
\bibinfo{author}{\bibfnamefont{N.~R.} \bibnamefont{{Tanvir}}},
  \bibinfo{author}{\bibfnamefont{A.~J.} \bibnamefont{{Levan}}},
  \bibinfo{author}{\bibfnamefont{A.~S.} \bibnamefont{{Fruchter}}},
  \bibinfo{author}{\bibfnamefont{J.}~\bibnamefont{{Hjorth}}},
  \bibinfo{author}{\bibfnamefont{K.}~\bibnamefont{{Wiersema}}},
  \bibinfo{author}{\bibfnamefont{R.}~\bibnamefont{{Tunnicliffe}}},
  \bibnamefont{and} \bibinfo{author}{\bibfnamefont{A.}~\bibnamefont{{de Ugarte
  Postigo}}}, \bibinfo{journal}{ArXiv e-prints}  (\bibinfo{year}{2013}),
  \eprint{1306.4971}.

\bibitem[{\citenamefont{{Shibata}}(2005)}]{2005PhRvL..94t1101S}
\bibinfo{author}{\bibfnamefont{M.}~\bibnamefont{{Shibata}}},
  \bibinfo{journal}{\prl} \textbf{\bibinfo{volume}{94}},
  \bibinfo{pages}{201101} (\bibinfo{year}{2005}).

\bibitem[{\citenamefont{{Shibata} et~al.}(2005)\citenamefont{{Shibata},
  {Taniguchi}, and {Ury{\= u}}}}]{2005PhRvD..71h4021S}
\bibinfo{author}{\bibfnamefont{M.}~\bibnamefont{{Shibata}}},
  \bibinfo{author}{\bibfnamefont{K.}~\bibnamefont{{Taniguchi}}},
  \bibnamefont{and} \bibinfo{author}{\bibfnamefont{K.}~\bibnamefont{{Ury{\=
  u}}}}, \bibinfo{journal}{\prd} \textbf{\bibinfo{volume}{71}},
  \bibinfo{eid}{084021} (\bibinfo{year}{2005}).

\bibitem[{\citenamefont{{Shibata} and {Taniguchi}}(2006)}]{2006PhRvD..73f4027S}
\bibinfo{author}{\bibfnamefont{M.}~\bibnamefont{{Shibata}}} \bibnamefont{and}
  \bibinfo{author}{\bibfnamefont{K.}~\bibnamefont{{Taniguchi}}},
  \bibinfo{journal}{\prd} \textbf{\bibinfo{volume}{73}}, \bibinfo{eid}{064027}
  (\bibinfo{year}{2006}).

\bibitem[{\citenamefont{{Oechslin} et~al.}(2007)\citenamefont{{Oechslin},
  {Janka}, and {Marek}}}]{2007A&A...467..395O}
\bibinfo{author}{\bibfnamefont{R.}~\bibnamefont{{Oechslin}}},
  \bibinfo{author}{\bibfnamefont{H.-T.} \bibnamefont{{Janka}}},
  \bibnamefont{and} \bibinfo{author}{\bibfnamefont{A.}~\bibnamefont{{Marek}}},
  \bibinfo{journal}{Astron. Astrophys.} \textbf{\bibinfo{volume}{467}},
  \bibinfo{pages}{395} (\bibinfo{year}{2007}).

\bibitem[{\citenamefont{{Anderson} et~al.}(2008)\citenamefont{{Anderson},
  {Hirschmann}, {Lehner}, {Liebling}, {Motl}, {Neilsen}, {Palenzuela}, and
  {Tohline}}}]{2008PhRvD..77b4006A}
\bibinfo{author}{\bibfnamefont{M.}~\bibnamefont{{Anderson}}},
  \bibinfo{author}{\bibfnamefont{E.~W.} \bibnamefont{{Hirschmann}}},
  \bibinfo{author}{\bibfnamefont{L.}~\bibnamefont{{Lehner}}},
  \bibinfo{author}{\bibfnamefont{S.~L.} \bibnamefont{{Liebling}}},
  \bibinfo{author}{\bibfnamefont{P.~M.} \bibnamefont{{Motl}}},
  \bibinfo{author}{\bibfnamefont{D.}~\bibnamefont{{Neilsen}}},
  \bibinfo{author}{\bibfnamefont{C.}~\bibnamefont{{Palenzuela}}},
  \bibnamefont{and} \bibinfo{author}{\bibfnamefont{J.~E.}
  \bibnamefont{{Tohline}}}, \bibinfo{journal}{\prd}
  \textbf{\bibinfo{volume}{77}}, \bibinfo{eid}{024006} (\bibinfo{year}{2008}).

\bibitem[{\citenamefont{{Liu} et~al.}(2008)\citenamefont{{Liu}, {Shapiro},
  {Etienne}, and {Taniguchi}}}]{2008PhRvD..78b4012L}
\bibinfo{author}{\bibfnamefont{Y.~T.} \bibnamefont{{Liu}}},
  \bibinfo{author}{\bibfnamefont{S.~L.} \bibnamefont{{Shapiro}}},
  \bibinfo{author}{\bibfnamefont{Z.~B.} \bibnamefont{{Etienne}}},
  \bibnamefont{and}
  \bibinfo{author}{\bibfnamefont{K.}~\bibnamefont{{Taniguchi}}},
  \bibinfo{journal}{\prd} \textbf{\bibinfo{volume}{78}}, \bibinfo{eid}{024012}
  (\bibinfo{year}{2008}).

\bibitem[{\citenamefont{{Baiotti} et~al.}(2008)\citenamefont{{Baiotti},
  {Giacomazzo}, and {Rezzolla}}}]{2008PhRvD..78h4033B}
\bibinfo{author}{\bibfnamefont{L.}~\bibnamefont{{Baiotti}}},
  \bibinfo{author}{\bibfnamefont{B.}~\bibnamefont{{Giacomazzo}}},
  \bibnamefont{and}
  \bibinfo{author}{\bibfnamefont{L.}~\bibnamefont{{Rezzolla}}},
  \bibinfo{journal}{\prd} \textbf{\bibinfo{volume}{78}}, \bibinfo{eid}{084033}
  (\bibinfo{year}{2008}).

\bibitem[{\citenamefont{{Kiuchi} et~al.}(2009)\citenamefont{{Kiuchi},
  {Sekiguchi}, {Shibata}, and {Taniguchi}}}]{2009PhRvD..80f4037K}
\bibinfo{author}{\bibfnamefont{K.}~\bibnamefont{{Kiuchi}}},
  \bibinfo{author}{\bibfnamefont{Y.}~\bibnamefont{{Sekiguchi}}},
  \bibinfo{author}{\bibfnamefont{M.}~\bibnamefont{{Shibata}}},
  \bibnamefont{and}
  \bibinfo{author}{\bibfnamefont{K.}~\bibnamefont{{Taniguchi}}},
  \bibinfo{journal}{\prd} \textbf{\bibinfo{volume}{80}}, \bibinfo{eid}{064037}
  (\bibinfo{year}{2009}).

\bibitem[{\citenamefont{{Hotokezaka} et~al.}(2011)\citenamefont{{Hotokezaka},
  {Kyutoku}, {Okawa}, {Shibata}, and {Kiuchi}}}]{2011PhRvD..83l4008H}
\bibinfo{author}{\bibfnamefont{K.}~\bibnamefont{{Hotokezaka}}},
  \bibinfo{author}{\bibfnamefont{K.}~\bibnamefont{{Kyutoku}}},
  \bibinfo{author}{\bibfnamefont{H.}~\bibnamefont{{Okawa}}},
  \bibinfo{author}{\bibfnamefont{M.}~\bibnamefont{{Shibata}}},
  \bibnamefont{and} \bibinfo{author}{\bibfnamefont{K.}~\bibnamefont{{Kiuchi}}},
  \bibinfo{journal}{\prd} \textbf{\bibinfo{volume}{83}}, \bibinfo{eid}{124008}
  (\bibinfo{year}{2011}).

\bibitem[{\citenamefont{{Sekiguchi} et~al.}(2011)\citenamefont{{Sekiguchi},
  {Kiuchi}, {Kyutoku}, and {Shibata}}}]{2011PhRvL.107e1102S}
\bibinfo{author}{\bibfnamefont{Y.}~\bibnamefont{{Sekiguchi}}},
  \bibinfo{author}{\bibfnamefont{K.}~\bibnamefont{{Kiuchi}}},
  \bibinfo{author}{\bibfnamefont{K.}~\bibnamefont{{Kyutoku}}},
  \bibnamefont{and}
  \bibinfo{author}{\bibfnamefont{M.}~\bibnamefont{{Shibata}}},
  \bibinfo{journal}{Physical Review Letters} \textbf{\bibinfo{volume}{107}},
  \bibinfo{eid}{051102} (\bibinfo{year}{2011}).

\bibitem[{\citenamefont{{Bernuzzi} et~al.}(2012)\citenamefont{{Bernuzzi},
  {Thierfelder}, and {Br{\"u}gmann}}}]{2012PhRvD..85j4030B}
\bibinfo{author}{\bibfnamefont{S.}~\bibnamefont{{Bernuzzi}}},
  \bibinfo{author}{\bibfnamefont{M.}~\bibnamefont{{Thierfelder}}},
  \bibnamefont{and}
  \bibinfo{author}{\bibfnamefont{B.}~\bibnamefont{{Br{\"u}gmann}}},
  \bibinfo{journal}{\prd} \textbf{\bibinfo{volume}{85}}, \bibinfo{eid}{104030}
  (\bibinfo{year}{2012}).

\bibitem[{\citenamefont{{Bauswein} et~al.}(2012)\citenamefont{{Bauswein},
  {Janka}, {Hebeler}, and {Schwenk}}}]{2012PhRvD..86f3001B}
\bibinfo{author}{\bibfnamefont{A.}~\bibnamefont{{Bauswein}}},
  \bibinfo{author}{\bibfnamefont{H.-T.} \bibnamefont{{Janka}}},
  \bibinfo{author}{\bibfnamefont{K.}~\bibnamefont{{Hebeler}}},
  \bibnamefont{and}
  \bibinfo{author}{\bibfnamefont{A.}~\bibnamefont{{Schwenk}}},
  \bibinfo{journal}{\prd} \textbf{\bibinfo{volume}{86}}, \bibinfo{eid}{063001}
  (\bibinfo{year}{2012}).

\bibitem[{\citenamefont{{Rosswog} et~al.}(2013)\citenamefont{{Rosswog},
  {Piran}, and {Nakar}}}]{2013MNRAS.430.2585R}
\bibinfo{author}{\bibfnamefont{S.}~\bibnamefont{{Rosswog}}},
  \bibinfo{author}{\bibfnamefont{T.}~\bibnamefont{{Piran}}}, \bibnamefont{and}
  \bibinfo{author}{\bibfnamefont{E.}~\bibnamefont{{Nakar}}},
  \bibinfo{journal}{Mon. Not. Roy. Astron. Soc.}
  \textbf{\bibinfo{volume}{430}}, \bibinfo{pages}{2585} (\bibinfo{year}{2013}).

\bibitem[{\citenamefont{{Bauswein} et~al.}(2013)\citenamefont{{Bauswein},
  {Goriely}, and {Janka}}}]{2013arXiv1302.6530B}
\bibinfo{author}{\bibfnamefont{A.}~\bibnamefont{{Bauswein}}},
  \bibinfo{author}{\bibfnamefont{S.}~\bibnamefont{{Goriely}}},
  \bibnamefont{and} \bibinfo{author}{\bibfnamefont{H.-T.}
  \bibnamefont{{Janka}}}, \bibinfo{journal}{ArXiv e-prints}
  (\bibinfo{year}{2013}), \eprint{1302.6530}.

\bibitem[{\citenamefont{{Duez}}(2010)}]{2010CQGra..27k4002D}
\bibinfo{author}{\bibfnamefont{M.~D.} \bibnamefont{{Duez}}},
  \bibinfo{journal}{Classical and Quantum Gravity}
  \textbf{\bibinfo{volume}{27}}, \bibinfo{eid}{114002} (\bibinfo{year}{2010}).

\bibitem[{\citenamefont{{Baumgarte} and {Shapiro}}(2010)}]{BaumgarteShapiro}
\bibinfo{author}{\bibfnamefont{T.~W.} \bibnamefont{{Baumgarte}}}
  \bibnamefont{and} \bibinfo{author}{\bibfnamefont{S.~L.}
  \bibnamefont{{Shapiro}}}, \emph{\bibinfo{title}{{Numerical relativity:
  Solving Einstein's Equations on the Computer}}}
  (\bibinfo{publisher}{Cambridge University Press, Cambridge},
  \bibinfo{year}{2010}).

\bibitem[{\citenamefont{{Faber} and {Rasio}}(2012)}]{2012LRR....15....8F}
\bibinfo{author}{\bibfnamefont{J.~A.} \bibnamefont{{Faber}}} \bibnamefont{and}
  \bibinfo{author}{\bibfnamefont{F.~A.} \bibnamefont{{Rasio}}},
  \bibinfo{journal}{Living Reviews in Relativity}
  \textbf{\bibinfo{volume}{15}}, \bibinfo{pages}{8} (\bibinfo{year}{2012}).

\bibitem[{\citenamefont{{Lattimer}}(2012)}]{2012ARNPS..62..485L}
\bibinfo{author}{\bibfnamefont{J.~M.} \bibnamefont{{Lattimer}}},
  \bibinfo{journal}{Annual Review of Nuclear and Particle Science}
  \textbf{\bibinfo{volume}{62}}, \bibinfo{pages}{485} (\bibinfo{year}{2012}),
  \eprint{1305.3510}.

\bibitem[{\citenamefont{{Baumgarte} et~al.}(2000)\citenamefont{{Baumgarte},
  {Shapiro}, and {Shibata}}}]{2000ApJ...528L..29B}
\bibinfo{author}{\bibfnamefont{T.~W.} \bibnamefont{{Baumgarte}}},
  \bibinfo{author}{\bibfnamefont{S.~L.} \bibnamefont{{Shapiro}}},
  \bibnamefont{and}
  \bibinfo{author}{\bibfnamefont{M.}~\bibnamefont{{Shibata}}},
  \bibinfo{journal}{Astrophys. J. Lett.} \textbf{\bibinfo{volume}{528}},
  \bibinfo{pages}{L29} (\bibinfo{year}{2000}).

\bibitem[{\citenamefont{{Kaplan} et~al.}(2013)\citenamefont{{Kaplan}, {Ott},
  {O'Connor}, {Kiuchi}, {Roberts}, and {Duez}}}]{2013arXiv1306.4034K}
\bibinfo{author}{\bibfnamefont{J.~D.} \bibnamefont{{Kaplan}}},
  \bibinfo{author}{\bibfnamefont{C.~D.} \bibnamefont{{Ott}}},
  \bibinfo{author}{\bibfnamefont{E.~P.} \bibnamefont{{O'Connor}}},
  \bibinfo{author}{\bibfnamefont{K.}~\bibnamefont{{Kiuchi}}},
  \bibinfo{author}{\bibfnamefont{L.}~\bibnamefont{{Roberts}}},
  \bibnamefont{and} \bibinfo{author}{\bibfnamefont{M.}~\bibnamefont{{Duez}}},
  \bibinfo{journal}{ArXiv e-prints}  (\bibinfo{year}{2013}),
  \eprint{1306.4034}.

\bibitem[{\citenamefont{{Tolman}}(1939)}]{1939PhRv...55..364T}
\bibinfo{author}{\bibfnamefont{R.~C.} \bibnamefont{{Tolman}}},
  \bibinfo{journal}{Phys. Rev.} \textbf{\bibinfo{volume}{55}},
  \bibinfo{pages}{364} (\bibinfo{year}{1939}).

\bibitem[{\citenamefont{{Oppenheimer} and
  {Volkoff}}(1939)}]{1939PhRv...55..374O}
\bibinfo{author}{\bibfnamefont{J.~R.} \bibnamefont{{Oppenheimer}}}
  \bibnamefont{and} \bibinfo{author}{\bibfnamefont{G.~M.}
  \bibnamefont{{Volkoff}}}, \bibinfo{journal}{Phys. Rev.}
  \textbf{\bibinfo{volume}{55}}, \bibinfo{pages}{374} (\bibinfo{year}{1939}).

\bibitem[{\citenamefont{{Oechslin} et~al.}(2002)\citenamefont{{Oechslin},
  {Rosswog}, and {Thielemann}}}]{2002PhRvD..65j3005O}
\bibinfo{author}{\bibfnamefont{R.}~\bibnamefont{{Oechslin}}},
  \bibinfo{author}{\bibfnamefont{S.}~\bibnamefont{{Rosswog}}},
  \bibnamefont{and} \bibinfo{author}{\bibfnamefont{F.-K.}
  \bibnamefont{{Thielemann}}}, \bibinfo{journal}{\prd}
  \textbf{\bibinfo{volume}{65}}, \bibinfo{pages}{103005}
  (\bibinfo{year}{2002}).

\bibitem[{\citenamefont{{Bauswein} et~al.}(2010)\citenamefont{{Bauswein},
  {Janka}, and {Oechslin}}}]{2010PhRvD..82h4043B}
\bibinfo{author}{\bibfnamefont{A.}~\bibnamefont{{Bauswein}}},
  \bibinfo{author}{\bibfnamefont{H.-T.} \bibnamefont{{Janka}}},
  \bibnamefont{and}
  \bibinfo{author}{\bibfnamefont{R.}~\bibnamefont{{Oechslin}}},
  \bibinfo{journal}{\prd} \textbf{\bibinfo{volume}{82}},
  \bibinfo{pages}{084043} (\bibinfo{year}{2010}).

\bibitem[{\citenamefont{{Demorest} et~al.}(2010)\citenamefont{{Demorest},
  {Pennucci}, {Ransom}, {Roberts}, and {Hessels}}}]{2010Natur.467.1081D}
\bibinfo{author}{\bibfnamefont{P.~B.} \bibnamefont{{Demorest}}},
  \bibinfo{author}{\bibfnamefont{T.}~\bibnamefont{{Pennucci}}},
  \bibinfo{author}{\bibfnamefont{S.~M.} \bibnamefont{{Ransom}}},
  \bibinfo{author}{\bibfnamefont{M.~S.~E.} \bibnamefont{{Roberts}}},
  \bibnamefont{and} \bibinfo{author}{\bibfnamefont{J.~W.~T.}
  \bibnamefont{{Hessels}}}, \bibinfo{journal}{\nat}
  \textbf{\bibinfo{volume}{467}}, \bibinfo{pages}{1081} (\bibinfo{year}{2010}).

\bibitem[{\citenamefont{Antoniadis et~al.}(2013)\citenamefont{Antoniadis,
  Freire, Wex, Tauris, Lynch, van Kerkwijk, Kramer, Bassa, Dhillon, Driebe
  et~al.}}]{Antoniadis26042013}
\bibinfo{author}{\bibfnamefont{J.}~\bibnamefont{Antoniadis}},
  \bibinfo{author}{\bibfnamefont{P.~C.~C.} \bibnamefont{Freire}},
  \bibinfo{author}{\bibfnamefont{N.}~\bibnamefont{Wex}},
  \bibinfo{author}{\bibfnamefont{T.~M.} \bibnamefont{Tauris}},
  \bibinfo{author}{\bibfnamefont{R.~S.} \bibnamefont{Lynch}},
  \bibinfo{author}{\bibfnamefont{M.~H.} \bibnamefont{van Kerkwijk}},
  \bibinfo{author}{\bibfnamefont{M.}~\bibnamefont{Kramer}},
  \bibinfo{author}{\bibfnamefont{C.}~\bibnamefont{Bassa}},
  \bibinfo{author}{\bibfnamefont{V.~S.} \bibnamefont{Dhillon}},
  \bibinfo{author}{\bibfnamefont{T.}~\bibnamefont{Driebe}},
  \bibnamefont{et~al.}, \bibinfo{journal}{Science}
  \textbf{\bibinfo{volume}{340}} (\bibinfo{year}{2013}).

\bibitem[{\citenamefont{{Bildsten} and {Cutler}}(1992)}]{1992ApJ...400..175B}
\bibinfo{author}{\bibfnamefont{L.}~\bibnamefont{{Bildsten}}} \bibnamefont{and}
  \bibinfo{author}{\bibfnamefont{C.}~\bibnamefont{{Cutler}}},
  \bibinfo{journal}{Astrophys. J.} \textbf{\bibinfo{volume}{400}},
  \bibinfo{pages}{175} (\bibinfo{year}{1992}).

\bibitem[{\citenamefont{{Kochanek}}(1992)}]{1992ApJ...398..234K}
\bibinfo{author}{\bibfnamefont{C.~S.} \bibnamefont{{Kochanek}}},
  \bibinfo{journal}{\apj} \textbf{\bibinfo{volume}{398}}, \bibinfo{pages}{234}
  (\bibinfo{year}{1992}).

\bibitem[{\citenamefont{{Lalazissis} et~al.}(1997)\citenamefont{{Lalazissis},
  {K{\"o}nig}, and {Ring}}}]{1997PhRvC..55..540L}
\bibinfo{author}{\bibfnamefont{G.~A.} \bibnamefont{{Lalazissis}}},
  \bibinfo{author}{\bibfnamefont{J.}~\bibnamefont{{K{\"o}nig}}},
  \bibnamefont{and} \bibinfo{author}{\bibfnamefont{P.}~\bibnamefont{{Ring}}},
  \bibinfo{journal}{\prc} \textbf{\bibinfo{volume}{55}}, \bibinfo{pages}{540}
  (\bibinfo{year}{1997}).

\bibitem[{\citenamefont{{Hempel} and
  {Schaffner-Bielich}}(2010)}]{2010NuPhA.837..210H}
\bibinfo{author}{\bibfnamefont{M.}~\bibnamefont{{Hempel}}} \bibnamefont{and}
  \bibinfo{author}{\bibfnamefont{J.}~\bibnamefont{{Schaffner-Bielich}}},
  \bibinfo{journal}{Nucl. Phys. A} \textbf{\bibinfo{volume}{837}},
  \bibinfo{pages}{210} (\bibinfo{year}{2010}).

\bibitem[{\citenamefont{{Shen} et~al.}(2011{\natexlab{a}})\citenamefont{{Shen},
  {Horowitz}, and {Teige}}}]{2011PhRvC..83c5802S}
\bibinfo{author}{\bibfnamefont{G.}~\bibnamefont{{Shen}}},
  \bibinfo{author}{\bibfnamefont{C.~J.} \bibnamefont{{Horowitz}}},
  \bibnamefont{and} \bibinfo{author}{\bibfnamefont{S.}~\bibnamefont{{Teige}}},
  \bibinfo{journal}{\prc} \textbf{\bibinfo{volume}{83}},
  \bibinfo{pages}{035802} (\bibinfo{year}{2011}{\natexlab{a}}).

\bibitem[{\citenamefont{{Lattimer} and {Swesty}}(1991)}]{1991NuPhA.535..331L}
\bibinfo{author}{\bibfnamefont{J.~M.} \bibnamefont{{Lattimer}}}
  \bibnamefont{and} \bibinfo{author}{\bibfnamefont{F.~D.}
  \bibnamefont{{Swesty}}}, \bibinfo{journal}{Nucl. Phys. A}
  \textbf{\bibinfo{volume}{535}}, \bibinfo{pages}{331} (\bibinfo{year}{1991}).

\bibitem[{\citenamefont{{Typel} et~al.}(2010)\citenamefont{{Typel},
  {R{\"o}pke}, {Kl{\"a}hn}, {Blaschke}, and {Wolter}}}]{2010PhRvC..81a5803T}
\bibinfo{author}{\bibfnamefont{S.}~\bibnamefont{{Typel}}},
  \bibinfo{author}{\bibfnamefont{G.}~\bibnamefont{{R{\"o}pke}}},
  \bibinfo{author}{\bibfnamefont{T.}~\bibnamefont{{Kl{\"a}hn}}},
  \bibinfo{author}{\bibfnamefont{D.}~\bibnamefont{{Blaschke}}},
  \bibnamefont{and} \bibinfo{author}{\bibfnamefont{H.~H.}
  \bibnamefont{{Wolter}}}, \bibinfo{journal}{\prc}
  \textbf{\bibinfo{volume}{81}}, \bibinfo{eid}{015803} (\bibinfo{year}{2010}).

\bibitem[{\citenamefont{{Shen} et~al.}(1998)\citenamefont{{Shen}, {Toki},
  {Oyamatsu}, and {Sumiyoshi}}}]{1998NuPhA.637..435S}
\bibinfo{author}{\bibfnamefont{H.}~\bibnamefont{{Shen}}},
  \bibinfo{author}{\bibfnamefont{H.}~\bibnamefont{{Toki}}},
  \bibinfo{author}{\bibfnamefont{K.}~\bibnamefont{{Oyamatsu}}},
  \bibnamefont{and}
  \bibinfo{author}{\bibfnamefont{K.}~\bibnamefont{{Sumiyoshi}}},
  \bibinfo{journal}{Nucl. Phys. A} \textbf{\bibinfo{volume}{637}},
  \bibinfo{pages}{435} (\bibinfo{year}{1998}).

\bibitem[{\citenamefont{{Sugahara} and {Toki}}(1994)}]{1994NuPhA.579..557S}
\bibinfo{author}{\bibfnamefont{Y.}~\bibnamefont{{Sugahara}}} \bibnamefont{and}
  \bibinfo{author}{\bibfnamefont{H.}~\bibnamefont{{Toki}}},
  \bibinfo{journal}{Nuclear Physics A} \textbf{\bibinfo{volume}{579}},
  \bibinfo{pages}{557} (\bibinfo{year}{1994}).

\bibitem[{\citenamefont{{Hempel} et~al.}(2012)\citenamefont{{Hempel},
  {Fischer}, {Schaffner-Bielich}, and
  {Liebend{\"o}rfer}}}]{2012ApJ...748...70H}
\bibinfo{author}{\bibfnamefont{M.}~\bibnamefont{{Hempel}}},
  \bibinfo{author}{\bibfnamefont{T.}~\bibnamefont{{Fischer}}},
  \bibinfo{author}{\bibfnamefont{J.}~\bibnamefont{{Schaffner-Bielich}}},
  \bibnamefont{and}
  \bibinfo{author}{\bibfnamefont{M.}~\bibnamefont{{Liebend{\"o}rfer}}},
  \bibinfo{journal}{\apj} \textbf{\bibinfo{volume}{748}}, \bibinfo{eid}{70}
  (\bibinfo{year}{2012}).

\bibitem[{\citenamefont{{Steiner} et~al.}(2012)\citenamefont{{Steiner},
  {Hempel}, and {Fischer}}}]{2012arXiv1207.2184S}
\bibinfo{author}{\bibfnamefont{A.~W.} \bibnamefont{{Steiner}}},
  \bibinfo{author}{\bibfnamefont{M.}~\bibnamefont{{Hempel}}}, \bibnamefont{and}
  \bibinfo{author}{\bibfnamefont{T.}~\bibnamefont{{Fischer}}},
  \bibinfo{journal}{ArXiv e-prints}  (\bibinfo{year}{2012}),
  \eprint{1207.2184}.

\bibitem[{\citenamefont{{Shen} et~al.}(2011{\natexlab{b}})\citenamefont{{Shen},
  {Horowitz}, and {O'Connor}}}]{2011arXiv1103.5174S}
\bibinfo{author}{\bibfnamefont{G.}~\bibnamefont{{Shen}}},
  \bibinfo{author}{\bibfnamefont{C.~J.} \bibnamefont{{Horowitz}}},
  \bibnamefont{and}
  \bibinfo{author}{\bibfnamefont{E.}~\bibnamefont{{O'Connor}}},
  \bibinfo{journal}{\prc} \textbf{\bibinfo{volume}{83}}, \bibinfo{eid}{065808}
  (\bibinfo{year}{2011}{\natexlab{b}}).

\bibitem[{\citenamefont{{Toki} et~al.}(1995)\citenamefont{{Toki}, {Hirata},
  {Sugahara}, {Sumiyoshi}, and {Tanihata}}}]{1995NuPhA.588..357T}
\bibinfo{author}{\bibfnamefont{H.}~\bibnamefont{{Toki}}},
  \bibinfo{author}{\bibfnamefont{D.}~\bibnamefont{{Hirata}}},
  \bibinfo{author}{\bibfnamefont{Y.}~\bibnamefont{{Sugahara}}},
  \bibinfo{author}{\bibfnamefont{K.}~\bibnamefont{{Sumiyoshi}}},
  \bibnamefont{and}
  \bibinfo{author}{\bibfnamefont{I.}~\bibnamefont{{Tanihata}}},
  \bibinfo{journal}{Nuclear Physics A} \textbf{\bibinfo{volume}{588}},
  \bibinfo{pages}{357} (\bibinfo{year}{1995}).

\bibitem[{\citenamefont{{Fattoyev} et~al.}(2010)\citenamefont{{Fattoyev},
  {Horowitz}, {Piekarewicz}, and {Shen}}}]{2010PhRvC..82e5803F}
\bibinfo{author}{\bibfnamefont{F.~J.} \bibnamefont{{Fattoyev}}},
  \bibinfo{author}{\bibfnamefont{C.~J.} \bibnamefont{{Horowitz}}},
  \bibinfo{author}{\bibfnamefont{J.}~\bibnamefont{{Piekarewicz}}},
  \bibnamefont{and} \bibinfo{author}{\bibfnamefont{G.}~\bibnamefont{{Shen}}},
  \bibinfo{journal}{\prc} \textbf{\bibinfo{volume}{82}}, \bibinfo{eid}{055803}
  (\bibinfo{year}{2010}).

\bibitem[{\citenamefont{{Paschalidis} et~al.}(2012)\citenamefont{{Paschalidis},
  {Etienne}, and {Shapiro}}}]{2012PhRvD..86f4032P}
\bibinfo{author}{\bibfnamefont{V.}~\bibnamefont{{Paschalidis}}},
  \bibinfo{author}{\bibfnamefont{Z.~B.} \bibnamefont{{Etienne}}},
  \bibnamefont{and} \bibinfo{author}{\bibfnamefont{S.~L.}
  \bibnamefont{{Shapiro}}}, \bibinfo{journal}{\prd}
  \textbf{\bibinfo{volume}{86}}, \bibinfo{eid}{064032} (\bibinfo{year}{2012}).

\bibitem[{\citenamefont{{Read} et~al.}(2013)\citenamefont{{Read}, {Baiotti},
  {Creighton}, {Friedman}, {Giacomazzo}, {Kyutoku}, {Markakis}, {Rezzolla},
  {Shibata}, and {Taniguchi}}}]{2013arXiv1306.4065R}
\bibinfo{author}{\bibfnamefont{J.~S.} \bibnamefont{{Read}}},
  \bibinfo{author}{\bibfnamefont{L.}~\bibnamefont{{Baiotti}}},
  \bibinfo{author}{\bibfnamefont{J.~D.~E.} \bibnamefont{{Creighton}}},
  \bibinfo{author}{\bibfnamefont{J.~L.} \bibnamefont{{Friedman}}},
  \bibinfo{author}{\bibfnamefont{B.}~\bibnamefont{{Giacomazzo}}},
  \bibinfo{author}{\bibfnamefont{K.}~\bibnamefont{{Kyutoku}}},
  \bibinfo{author}{\bibfnamefont{C.}~\bibnamefont{{Markakis}}},
  \bibinfo{author}{\bibfnamefont{L.}~\bibnamefont{{Rezzolla}}},
  \bibinfo{author}{\bibfnamefont{M.}~\bibnamefont{{Shibata}}},
  \bibnamefont{and}
  \bibinfo{author}{\bibfnamefont{K.}~\bibnamefont{{Taniguchi}}},
  \bibinfo{journal}{ArXiv e-prints}  (\bibinfo{year}{2013}),
  \eprint{1306.4065}.

\bibitem[{\citenamefont{{Bauswein} and {Janka}}(2012)}]{2012PhRvL.108a1101B}
\bibinfo{author}{\bibfnamefont{A.}~\bibnamefont{{Bauswein}}} \bibnamefont{and}
  \bibinfo{author}{\bibfnamefont{H.-T.} \bibnamefont{{Janka}}},
  \bibinfo{journal}{\prl} \textbf{\bibinfo{volume}{108}}, \bibinfo{eid}{011101}
  (\bibinfo{year}{2012}).

\bibitem[{\citenamefont{{Hebeler} et~al.}(2010)\citenamefont{{Hebeler},
  {Lattimer}, {Pethick}, and {Schwenk}}}]{2010PhRvL.105p1102H}
\bibinfo{author}{\bibfnamefont{K.}~\bibnamefont{{Hebeler}}},
  \bibinfo{author}{\bibfnamefont{J.~M.} \bibnamefont{{Lattimer}}},
  \bibinfo{author}{\bibfnamefont{C.~J.} \bibnamefont{{Pethick}}},
  \bibnamefont{and}
  \bibinfo{author}{\bibfnamefont{A.}~\bibnamefont{{Schwenk}}},
  \bibinfo{journal}{\prl} \textbf{\bibinfo{volume}{105}},
  \bibinfo{pages}{161102} (\bibinfo{year}{2010}).

\bibitem[{\citenamefont{{Koranda} et~al.}(1997)\citenamefont{{Koranda},
  {Stergioulas}, and {Friedman}}}]{1997ApJ...488..799K}
\bibinfo{author}{\bibfnamefont{S.}~\bibnamefont{{Koranda}}},
  \bibinfo{author}{\bibfnamefont{N.}~\bibnamefont{{Stergioulas}}},
  \bibnamefont{and} \bibinfo{author}{\bibfnamefont{J.~L.}
  \bibnamefont{{Friedman}}}, \bibinfo{journal}{\apj}
  \textbf{\bibinfo{volume}{488}}, \bibinfo{pages}{799} (\bibinfo{year}{1997}).

\bibitem[{\citenamefont{{Lyford} et~al.}(2003)\citenamefont{{Lyford},
  {Baumgarte}, and {Shapiro}}}]{2003ApJ...583..410L}
\bibinfo{author}{\bibfnamefont{N.~D.} \bibnamefont{{Lyford}}},
  \bibinfo{author}{\bibfnamefont{T.~W.} \bibnamefont{{Baumgarte}}},
  \bibnamefont{and} \bibinfo{author}{\bibfnamefont{S.~L.}
  \bibnamefont{{Shapiro}}}, \bibinfo{journal}{\apj}
  \textbf{\bibinfo{volume}{583}}, \bibinfo{pages}{410} (\bibinfo{year}{2003}).

\bibitem[{\citenamefont{{Lai} et~al.}(1993)\citenamefont{{Lai}, {Rasio}, and
  {Shapiro}}}]{1993ApJS...88..205L}
\bibinfo{author}{\bibfnamefont{D.}~\bibnamefont{{Lai}}},
  \bibinfo{author}{\bibfnamefont{F.~A.} \bibnamefont{{Rasio}}},
  \bibnamefont{and} \bibinfo{author}{\bibfnamefont{S.~L.}
  \bibnamefont{{Shapiro}}}, \bibinfo{journal}{Astrophys. J. Suppl. Ser.}
  \textbf{\bibinfo{volume}{88}}, \bibinfo{pages}{205} (\bibinfo{year}{1993}).

\bibitem[{\citenamefont{{Finn} and {Chernoff}}(1993)}]{1993PhRvD..47.2198F}
\bibinfo{author}{\bibfnamefont{L.~S.} \bibnamefont{{Finn}}} \bibnamefont{and}
  \bibinfo{author}{\bibfnamefont{D.~F.} \bibnamefont{{Chernoff}}},
  \bibinfo{journal}{\prd} \textbf{\bibinfo{volume}{47}}, \bibinfo{pages}{2198}
  (\bibinfo{year}{1993}).

\bibitem[{\citenamefont{{Cutler} and {Flanagan}}(1994)}]{1994PhRvD..49.2658C}
\bibinfo{author}{\bibfnamefont{C.}~\bibnamefont{{Cutler}}} \bibnamefont{and}
  \bibinfo{author}{\bibfnamefont{{\'E}.~E.} \bibnamefont{{Flanagan}}},
  \bibinfo{journal}{\prd} \textbf{\bibinfo{volume}{49}}, \bibinfo{pages}{2658}
  (\bibinfo{year}{1994}).

\bibitem[{\citenamefont{{Arun} et~al.}(2005)\citenamefont{{Arun}, {Iyer},
  {Sathyaprakash}, and {Sundararajan}}}]{2005PhRvD..71h4008A}
\bibinfo{author}{\bibfnamefont{K.~G.} \bibnamefont{{Arun}}},
  \bibinfo{author}{\bibfnamefont{B.~R.} \bibnamefont{{Iyer}}},
  \bibinfo{author}{\bibfnamefont{B.~S.} \bibnamefont{{Sathyaprakash}}},
  \bibnamefont{and} \bibinfo{author}{\bibfnamefont{P.~A.}
  \bibnamefont{{Sundararajan}}}, \bibinfo{journal}{\prd}
  \textbf{\bibinfo{volume}{71}}, \bibinfo{eid}{084008} (\bibinfo{year}{2005}).

\bibitem[{\citenamefont{{Hannam} et~al.}(2013)\citenamefont{{Hannam}, {Brown},
  {Fairhurst}, {Fryer}, and {Harry}}}]{2013ApJ...766L..14H}
\bibinfo{author}{\bibfnamefont{M.}~\bibnamefont{{Hannam}}},
  \bibinfo{author}{\bibfnamefont{D.~A.} \bibnamefont{{Brown}}},
  \bibinfo{author}{\bibfnamefont{S.}~\bibnamefont{{Fairhurst}}},
  \bibinfo{author}{\bibfnamefont{C.~L.} \bibnamefont{{Fryer}}},
  \bibnamefont{and} \bibinfo{author}{\bibfnamefont{I.~W.}
  \bibnamefont{{Harry}}}, \bibinfo{journal}{Astrophys. J. Lett.}
  \textbf{\bibinfo{volume}{766}}, \bibinfo{eid}{L14} (\bibinfo{year}{2013}).

\bibitem[{\citenamefont{{The LIGO Scientific Collaboration}
  et~al.}(2013)\citenamefont{{The LIGO Scientific Collaboration}, {the Virgo
  Collaboration}, {Aasi}, {Abadie}, {Abbott}, {Abbott}, {Abbott}, {Abernathy},
  {Accadia}, {Acernese} et~al.}}]{2013arXiv1304.1775T}
\bibinfo{author}{\bibnamefont{{The LIGO Scientific Collaboration}}},
  \bibinfo{author}{\bibnamefont{{the Virgo Collaboration}}},
  \bibinfo{author}{\bibfnamefont{J.}~\bibnamefont{{Aasi}}},
  \bibinfo{author}{\bibfnamefont{J.}~\bibnamefont{{Abadie}}},
  \bibinfo{author}{\bibfnamefont{B.~P.} \bibnamefont{{Abbott}}},
  \bibinfo{author}{\bibfnamefont{R.}~\bibnamefont{{Abbott}}},
  \bibinfo{author}{\bibfnamefont{T.~D.} \bibnamefont{{Abbott}}},
  \bibinfo{author}{\bibfnamefont{M.}~\bibnamefont{{Abernathy}}},
  \bibinfo{author}{\bibfnamefont{T.}~\bibnamefont{{Accadia}}},
  \bibinfo{author}{\bibfnamefont{F.}~\bibnamefont{{Acernese}}},
  \bibnamefont{et~al.}, \bibinfo{journal}{ArXiv e-prints}
  (\bibinfo{year}{2013}), \eprint{1304.1775}.

\end{thebibliography}

\end{document}